# Pattern of Impact-Induced Ejecta from Granular Targets with Large Inclusions

Toshihiko Kadono[1], Ryo Suetsugu[1], Dai Arakawa[1], Yoshiki Kasagi[1], Syuichi Nagayama[1], Ayako I. Suzuki[2], and Sunao Hasegawa[2]

[1] Department of Basic Sciences, University of Occupational and Environmental Health, Yahata, Kitakyusyu, Japan

[2] Institute of Space and Astronautical Science, Japan Aerospace Exploration Agency, Sagamihara, Kanagawa, Japan

\* Corresponding author.

*E-mail address:* kadono@med.uoeh-u.ac.jp (T. Kadono)

**Abstract**

We performed impact experiments to observe patterns in an ejecta curtain with targets consisting of small sand particles and large inclusions comparable to or smaller than the size of the projectiles. The spatial intensity distributions in the ejecta at early stages of crater formation depend on the size of the inclusions. Our numerical simulations of radially spreading particles with different sizes support this result. Based on the results, we proposed a procedure for evaluating the subsurface structures of celestial bodies from the images of ejecta curtains obtained from space-impact experiments.

## 1. Introduction

JAXA's planetary spacecraft, Hayabusa2, has arrived at the C-type asteroid Ryugu. Remote-sensing measurements have been performed since the arrival of the spacecraft, and various data about the asteroid have been obtained (e.g., Watanabe et al. 2019; Sugita et al. 2019). In addition to the remote-sensing instruments, Hayabusa2 carries an instrument called the Small Carry-on Impactor (SCI), which consists of a spherical copper shell projectile resembling a hollow ball that measures 13 cm in a diameter, with a mass of 2 kg and a bulk density of 2.3 g cm$^{-3}$ at an impact velocity of 2 km s$^{-1}$ (Arakawa et al. 2017). A key purpose of this instrument is to perform space-impact experiments on real asteroid materials at real asteroid scales under microgravity. Such space-impact experiments have been successfully performed on a comet in the DEEP IMPACT mission (e.g., A'Hearn et al. 2005) and on the moon in the LCROSS mission (e.g., Colaprete et al. 2010). The surface and subsurface structures of these bodies have since been assessed (e.g., Kadono et al. (2007) and Schultz et al. (2010), respectively). Meanwhile, the space experiment using SCI is the first ever impact experiment on an asteroid. One advantage of the Hayabusa2 for space-impact experiments is the use of the Deployable CAMera-3 (DCAM3), which is released from the spacecraft to observe the cratering process at a distance of 1 km from the impact point with a resolution less than 1 m/pixel and field of view of 74° (Ogawa et al. 2017). This feature allows the formation of the ejecta curtain to be observed in situ.

In this study, we proposed a procedure for obtaining information about the surface geology and subsurface structure around the impact point on Ryugu using the observational results of the ejecta curtain with the DCAM3. We focused on the pattern observed in ejecta curtains. In laboratories, there are some studies on the pattern in ejecta curtains; Arakawa et al. (2017) have investigated the morphology of ejecta curtains caused by cratering on a single block, and Kadono et al. (2015) have investigated the pattern in ejecta curtains using targets consisting of identical fine sands. However, detailed global imaging observations of Ryugu have revealed that its surface is covered in boulders with a power-law size distribution (e.g., Sugita et al. 2019). Therefore, we investigated the effects of particle size distributions in targets on the pattern observed in ejecta curtains, which has not yet been evaluated in detail. In particular, we considered the case where the target consists of fine sands and larger particles comparable in size to or slightly smaller than the projectile to represent an SCI impact on a fine-grained layer including pebbles (5–15 cm) or coarse-grains (1 mm–5 cm). In order to properly replicate the surface conditions of Ryugu in the laboratory, using a 4.8-mm projectile at a nominal velocity of 2.5 km s$^{-1}$, we performed impact experiments with granular targets of 0.1-mm glass beads as a fine sand matrix mixed with 1-mm and/or 4-mm glass spheres as pebbles or coarse-grains. For various mixing ratios in weight % (wt%), we observed the patterns in the ejecta curtain reflecting the characteristics of the targets and compared these results to numerical simulations of radially spreading particles with different sizes; we then discussed the effects of inclusions on the pattern. Finally, we proposed a procedure for evaluating the subsurface structures of Ryugu based on the images obtained using the DCAM3.

The remainder of this paper includes a section detailing the experimental and numerical simulation methods, followed by the sections of the results and discussion.

## 2. Methods

### *2.1. Experiments*

We performed impact experiments using a two-stage hydrogen-gas gun at the Institute of Space and Astronautical Science, JAXA. Spherical polycarbonate projectiles with a diameter of 4.8 mm (0.068 g in mass) were accelerated to a nominal velocity of 2.5 km s$^{-1}$ to perpendicularly impact the surfaces of granular targets. Three types of experiments (a total of eight shots) were performed. The first series focused on changing the mixing ratio of 1-mm glass spheres to 0.1-mm glass beads in wt%, from 20:80 wt%, to 50:50 wt%, to 66:34 wt%, and finally to 100:0 wt%. The second series involved changing the mixing ratio of 4-mm glass spheres to 0.1-mm glass beads in wt%, from 20:80 wt%, to 50:50 wt%, and finally to 66:34 wt%. The third series included mixing 1-mm spheres, 4-mm spheres, and 0.1-mm beads at a ratio of 20:20:60 wt%. These targets were poured into a bowl, which had a radius of 15 cm and depth of 10 cm with a flat bottom, and set in a vacuum chamber. The ambient pressure was less than 2.0 Pa. The experimental conditions are summarized in Table 1. The ejecta motion was then observed using a high-speed video camera (HPV-X, Shimadzu Co. Ltd) with a framing speed of 500 frames per second. Two lights were set outside the chamber illuminating the ejecta from both sides.

**Table 1**

Experimental Conditions

| Shot No. | Inclusion | Inclusion content (wt %) | Impact velocity (km s$^{-1}$) |
|---|---|---|---|
| 420 | 1 mm | 20 | 2.51 |
| 422 | 1 mm | 50 | 2.58 |
| 424 | 1 mm | 66 | 2.57 |
| 425 | 1 mm | 100 | 2.62 |
| 421 | 4 mm | 20 | 2.49 |
| 423 | 4 mm | 50 | 2.54 |
| 428 | 4 mm | 66 | 2.69 |
| 429 | 1 mm + 4 mm | 20 + 20 | 2.68 |

*2.2. Numerical Simulations*

We numerically investigated the pattern formation of radially spreading inelastic particles with different sizes using the open source discrete element method simulator LIGGGHTS (e.g., Kloss et al. 2012) in which the individual particles are soft spheres and the interactions between particles in contact are taken into account. The normal repulsive forces were represented by a spring and dashpot in parallel. When two particles collided, the normal velocity component, $v_n$, between them was set to $-ev_n$, where $e$ is the coefficient of restitution and set to 0.1. The parameter characterizing the friction force acting between particles was fixed to 0.05. The particle density, Young modulus, and Poisson ratio, were 2.5 g cm$^{-3}$, 5 MPa, and 0.45, respectively. The gravitational force was included. The gravitational acceleration $g$ was set to 9.81 m s$^{-2}$.

However, the cohesive forces were not included for simplicity.

We considered targets consisting of spherical particles with a radius of 0.5 mm as a matrix and larger particles with radii of 1 mm, 2.5 mm, and 5 mm as inclusions. For each inclusion size, we set two mixing ratios of 80:20 and 60:40; hence, six runs were simulated in total. We prepared $10^5$ particles, and the entire computational domain had a 0.8-m × 0.8-m rectangular shape with a height of 0.5 m. Initially, the particles were accumulated in a hypothetical cylinder with a radius of 0.05 m and height of 0.15 m. Then, the cylinder collided vertically at the center of the bottom plate with a velocity of 20 m s$^{-1}$. There was no wall except the bottom surface, and the particles were removed when they reached the side boundaries.

## 3. Results

### 3.1. Experimental Results

#### 3.1.1 Pattern in the Ejecta

Figure 1 depicts two snapshots of the ejecta 40 ms after impact for the targets, which include (a) 1-mm spheres with a content of 20 wt% and (b) 4-mm spheres with a content of 20 wt%. Projectile coming from the top impacted perpendicular to the surface of the granular target in both cases. A mesh pattern appears in Fig. 1(a), whereas a filament pattern is presented in Fig. 1(b). Kadono et al. (2015) have also observed the mesh pattern for a target with only 0.1-mm beads. However, the spaces in the pattern in Fig. 1(a) are slightly larger.

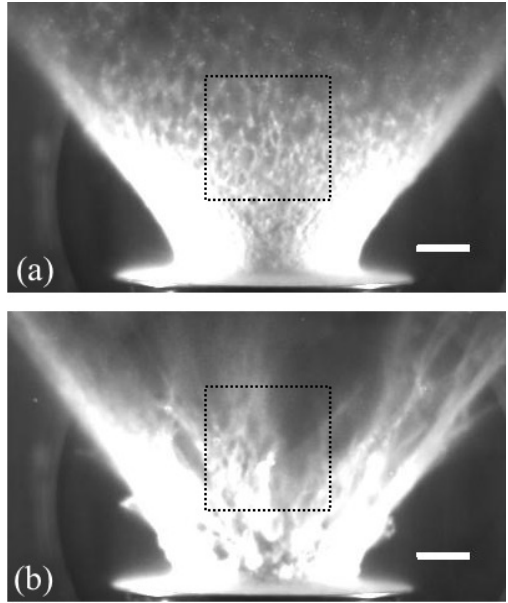

**Figure 1.** Snapshots of the ejecta for (a) the 1-mm glass sphere (20 wt%) + 0.1-mm glass bead target and (b) the 4-mm glass sphere (20 wt%) + 0.1-mm glass bead target 40 ms after impact. The white horizontal line in each frame is a spatial scale indicating 50 mm. To evaluate the spatial concentration of particles, we investigated the intensity in the area indicated by a square in each figure.

*3.1.2. Concentration of Particles in the Ejecta*

The mesh and filament patterns in Fig. 1 indicate that the granular media in the ejecta are spatially highly concentrated and that the spatial distributions of the particles are non-uniform. Herein, we investigate the intensity within an area of $N_0 = 100 \times 100$ pixels around the center of the ejecta curtain examples of which are shown in Figs. 1(a) and 1(b). Figure 2(a) illustrates the intensity distributions for the targets including the 1-mm spheres at 20 wt% (thin black curve) and the 4-mm spheres at 20 wt% (bold red curve) 40 ms after the impact. The spikes at the largest intensity indicate the saturation point. It appears that the contrast for the case with 4-mm inclusions is higher. Figure 2(b) presents the cumulative number of the pixels with intensities smaller than $I$, $N(< I)$ —using the same data as Fig. 2(a)—which is normalized according to $N_0$ (=$10^4$). As an index of the intensity contrast, we considered the ratios of the intensities at $N(< I)/N_0$ of 0.9 and 0.1 (denoted $I_{90}/I_{10}$). When this value is high, the intensity contrast and the particle concentration are high, and when it is low (~1), the particles are uniformly distributed. Figure 2(c) shows $I_{90}/I_{10}$ for each shot as a function of the normalized time, $\tau$, which is the time after the impact, $t$, divided by a characteristic crater formation time scale, $t_0$. We evaluated $t_0$ to be ~$(D_c/g)^{1/2}$ (Melosh 1989), where $D_c$ is the crater diameter (~10–20 cm for our impact conditions) and $g$ is the gravitational acceleration. We obtained $t_0$ to be ~100 ms in our cases; therefore, we set $t_0$ to 100 ms. For comparison purposes, the result of the target without inclusions (only 0.1-mm beads) is also shown. For the targets with 4-mm spheres, $I_{90}/I_{10}$ at early stages is extremely large (>~2.5), and it scatters and then rapidly decreases. Conversely, $I_{90}/I_{10}$ for the 1-mm targets is ~2–2.5 at early stages; then it approaches the curve for the target without inclusions. Meanwhile, the value of $I_{90}/I_{10}$ for the case with only 1-mm inclusions slowly decreases. By contrast, the case of the target that includes both 4-mm and 1-mm spheres appears to have intermediate behavior.

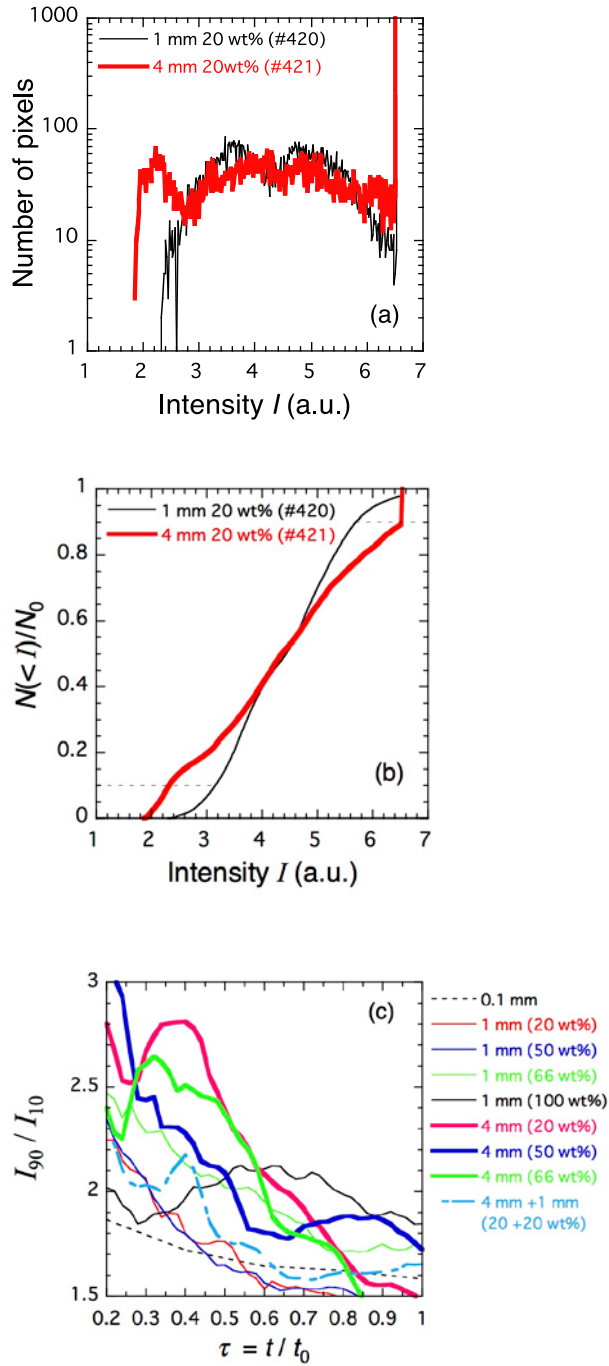

**Figure 2.** (a) Intensity distributions 40 ms after impact within the area with $N_0 = 100 \times 100$ pixels: the target including 1-mm spheres at 20 wt% (#420) (thin black curve) and the target including 4-mm spheres at 20 wt% (#421) (bold red curve). (b) Cumulative number of pixels having an intensity lower than $I$, $N(<I)$, normalized by $N_0$: the target

that includes 1-mm spheres at 20 wt% (#420) (thin black curve) and the target that includes 4-mm spheres at 20 wt% (#421) (bold red curve). The horizontal broken lines indicate $N(<I)/N_0$ of 0.9 and 0.1. (c) The ratio of the intensities at $N(<I)/N_0$ of 0.9 and 0.1 as a function of the normalized time, $\tau$, which is the time after the impact, $t$, divided by a characteristic crater formation time, $t_0$, which we set to 100 ms.

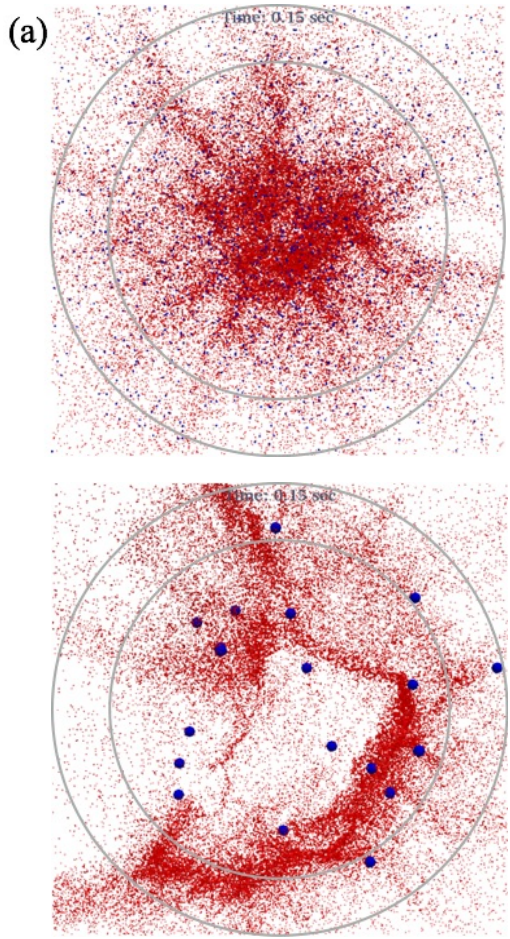

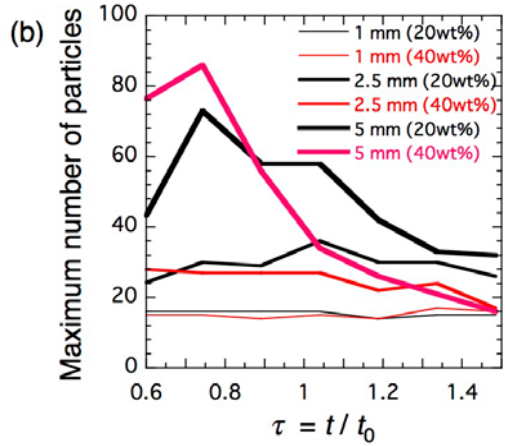

**Figure 3.** (a) Top view of the snapshots 0.15 s after impact in the numerical simulation (corresponding to the normalized time $\tau = 0.66$): 1-mm inclusions at 20 wt% (upper) and 5-mm inclusions at 20 wt% (lower). Small and large particles are represented by red and blue points, respectively. The particle sizes in the panels are exaggerated for visibility. The two concentric circles in each snapshot indicate the radii of 0.3 and 0.4 m. (b) We divide the annulus with radii between 0.3 and 0.4 m into 2560 boxes and consider the maximum number of small particles per box plotted as a function of the normalized time, which is the time after the impact, $t$, divided by the characteristic time scale, $t_0$. The concentration is high when the targets include large inclusions.

*3.2. Results of Numerical Simulations*

Figure 3(a) displays the top view of the patterns of radially spreading particles 0.15 s after impact: 1-mm inclusions at 20 wt% (upper) and 5-mm inclusions at 20 wt% (lower). We considered an annulus with radii between 0.3 and 0.4 m, which is indicated by the two concentric circles in Fig. 3(a). We divided this annulus into 2560 boxes (256 in the azimuth direction × 10 in the radial direction), and the number of particles in each box was counted. To evaluate the particle concentration, as shown in Fig. 3(b), the largest number of particles per box is shown as a function of the normalized time, $\tau$, for the 1-mm (thin curves), 2.5-mm (intermediate curves), and 5-mm (bold curves) inclusions with mixing ratios of 20 wt% (black curves) and 40 wt% (red curves). The normalized time, $\tau$, is the time after the impact, $t$, divided by the characteristic time scale, $t_0$, defined as $(r/g)^{1/2}$, where $r$ is set to 0.4 m. This figure corresponds to Fig. 2(c) in the experiments. It appears that the concentration increases with the size of the inclusions at early stages. This observation is consistent with the experimental results.

## 4. Discussion

### 4.1. Pattern Formation Process

As the mechanism for pattern formation, Kadono et al. (2015) have proposed the mutual inelastic collision of particles with fluctuating velocities during excavation when the targets consist of identical particles. In the case of targets that include large-sized particles, two processes are expected. One is mutual collision. When larger particles are included, coalescence should be promoted because their collision cross-section is larger. The other expected process is perturbation by large inclusions as obstacles. Various flow patterns are caused by inclusions, such as drags and spurts, which are typical in the experimental cases with targets that include 4-mm spheres and in the simulated cases with 5-mm particles.

### 4.2. Application to Space Impact Experiments

Based on our experimental and simulated results, we proposed a procedure for analyzing photos of ejecta curtains taken in space-impact experiments. First, in the spatial intensity distribution in the ejecta curtain as shown in Fig. 2(a), $N(<I)/N_0$ must be plotted as in Fig. 2(b). In the ratio of the intensities at $N(<I)/N_0$ of 0.9 and 0.1, $I_{90}/I_{10}$ must then be evaluated as a function of time as in Fig. 2(c). The time should be normalized by the crater formation time defined as the square root of the crater diameter over the gravitational acceleration. The crater diameter and gravitational acceleration can be obtained based on the remote-sensing measurements. The ratio $I_{90}/I_{10}$ at early stages represents the size of the pebbles or coarse-grains as inclusions in fine particles; when the pebble or coarse-grain size is ≥10 times larger than the size of the fine matrix particles, $I_{90}/I_{10}$ is high (>2.5) and scatters largely. Conversely, when the pebble or coarse-grain size is less than 10, $I_{90}/I_{10}$ is intermediate (~2–2.5) and gently decreases. When pebbles or coarse-grains are not included (only identical fine particles) or only identical pebbles or coarse-grains are included (no fine particles), $I_{90}/I_{10}$ is small (~1.5–2) and nearly constant. If the SCI projectile collides with a boulder field or a single block, the clear pattern would not be recognized in the ejecta.

Thus, because the intensity distribution in the images of the ejecta (i.e., particle concentration in the ejecta) depends on the size of inclusions, we presented a procedure to estimate the size ratio of inclusions and fine matrix particles. Although the accuracy of the estimation is an order of magnitude at present, using the procedure we can clarify whether the particles from the subsurface of Ryugu observed in the ejecta have order of magnitude differences in sizes or identical sizes. Such elucidation of the size distribution of the particles in the subsurface layer would lead us to discuss the formation process of Ryugu.

## 5. Summary

We experimentally investigated the patterns observed in impact-induced ejecta with targets consisting of small matrix sand particles and larger inclusions. Large inclusions promote pattern formation and perturb the excavation flow of the small sand particles. Therefore, the spatial intensity distribution in the ejecta curtain at early stages of crater formation depends on the size of the inclusions. Our numerical simulations support this result. Based on these observations, we proposed a procedure for evaluating the subsurface structures of Ryugu from the images obtained in the SCI experiment.

The authors wish to thank M. Arakawa, K. Ogawa, and K. Wada for helpful comments, M. Koga for supporting the data analysis, and M. Kiuchi and K. Ishiyama for supporting the impact experiments. The authors are also grateful to an anonymous reviewer for useful comments. This work was supported by ISAS/JAXA as a collaborative program with the Hypervelocity Impact Facility.